\documentclass[twocolumn,showpacs,preprintnumbers,amsmath,amssymb]{revtex4}

\usepackage{graphicx}
\usepackage{dcolumn}
\usepackage{bm}
\usepackage{graphicx}
\usepackage{epsfig}
\usepackage{rotating}

\begin{document}

\title{\boldmath Search for the decays
                 $J/\psi\to\gamma \rho \phi$ and
                 $J/\psi\to\gamma \rho \omega$}
\author{
M.~Ablikim$^{1}$, J.~Z.~Bai$^{1}$, Y.~Ban$^{12}$, X.~Cai$^{1}$,
H.~F.~Chen$^{17}$, H.~S.~Chen$^{1}$, H.~X.~Chen$^{1}$,
J.~C.~Chen$^{1}$, Jin~Chen$^{1}$, Y.~B.~Chen$^{1}$
% ...
\\(BES Collaboration)\\
$^{1}$ Institute of High Energy Physics, Beijing 100049, People's Republic of China\\
% ...
$^{12}$ Peking University, Beijing 100871, People's Republic of China\\
% ...
$^{17}$ University of Science and Technology of China, Hefei 230026, People's Republic of China\\
% ...
}
\date{\today}
\begin{abstract}
Using 58 million $J/\psi$ events collected with the Beijing
Spectrometer (BESII) at the Beijing Electron-Positron Collider, the
decays  $J/\psi\to \gamma\phi\rho$ and $J/\psi\to \gamma\omega\rho$  are
searched for, and upper limits on their branching fractions are
reported at the 90\%~C.~L. No clear structures are observed in the $\gamma
\rho$, $\gamma \phi$, or $\rho \phi $ mass spectra for $J/\psi\to
\gamma\phi\rho$ nor in the $\gamma \rho$, $\gamma \omega$, or $\rho
\omega$ mass spectra for $J/\psi\to \gamma\omega\rho$.
\end{abstract}


\maketitle
\section{Introduction}\label{int}
QCD predicts a rich spectrum of $gg$ glueballs, $ggq$ hybrids and
$qq\bar{q}\bar{q}$ four quark states along with the ordinary
$q\bar{q}$ mesons in the 1.0 to 2.5 $\hbox{GeV}/c^2$ mass region.
Radiative $J/\psi$ decays provide an excellent laboratory to search
for these states. Until now, no unique experimental signatures of
such states have been found.

Systems of two vector particles have been intensively studied for
signatures of gluonic bound states. Pseudoscalar ($0^{-}$)
enhancements in $\rho\rho$, $\omega\omega$, and $\phi\phi$ final
states have been seen in radiative $J/\psi$
decays~\cite{gvv,bal,bur,bal1,bai,bis1,bis2}, and a scalar ($0^{+}$)
enhancement near $\omega\phi$ threshold is observed from the doubly
OZI suppressed decay of $J/\psi\to \gamma \omega\phi$ with mass
$M=1812^{+19}_{-26}\pm18$ $\hbox{MeV}/c^2$ and width
$\Gamma=105\pm20\pm28$ $\hbox{MeV}/c^2$~\cite{pwa}. The radiative
$J/\psi$ decays $J/\psi\to\gamma \rho \phi$ and $J/\psi\to\gamma
\rho \omega$ are OZI suppressed processes, and the measurements of these
two decays and the search for possible resonant states in their
decay products will provide useful information on two vector meson
systems.

The double radiative channels $J/\psi\to \gamma X$, $X\to \gamma V$
($V$=$\rho$, $\phi$ and $\omega$) are studied to probe the quark
content of the object $X$. The $\eta(1440)$ has been studied by the
BES Collaboration through the double radiative channels
$J/\psi\to\gamma(\gamma\rho)$ and $J/\psi\to\gamma(\gamma\phi)$
~\cite{pid}. At one time, the $\eta(1440)$, after it was observed in
$J/\psi$ decay~\cite{1440}, was regarded as a glueball candidate. But
this viewpoint changed when its radiative decay modes were
observed. The $\eta(1440)$ is seen (at 1424 $\hbox{MeV}/c^2$) by the BES
Collaboration to decay strongly into $\gamma \rho$, not $\gamma
\phi$. From this result, one cannot draw a definite conclusion on
whether the $\eta(1440)$ is either a $q\bar{q}$ state or a glueball
state. Therefore, further study is needed to clarify the
situation. The process $J/\psi\to V X, X \to \gamma V$ also allows us
to study $X$ using the $\gamma V$ system.

In this letter, we report on the measurements of $J/\psi\to\gamma
\rho \phi$ and $J/\psi\to\gamma \rho \omega$ decays and the search
for possible structure in the $\gamma V$ and $V V$ invariant mass
spectra, using 58M $J/\psi$ events collected with the Beijing
Spectrometer (BESII) at the Beijing Electron-Position Collider
(BEPC)~\cite{bes}.

\section{Detector and Data Analysis}
BESII is a conventional solenoidal magnet detector that is described
in detail in Refs.~\cite{bes}. A 12-layer vertex chamber (VC)
surrounding the beam pipe provides coordinate and trigger
information. A forty-layer main drift chamber (MDC), located
radially outside the VC, provides trajectory and energy loss
($dE/dx$) information for tracks over 85\% of the total solid angle.
The momentum resolution is $\sigma _p/p = 0.017\sqrt{1+p^2}$ ~($p$
in GeV/$c$), and the $dE/dx$ resolution for hadron tracks is
$\sim$8\%. An array of 48 scintillation counters surrounding the MDC
measures the time-of-flight (TOF) of tracks with a resolution of
$\sim$200 ps for hadrons.  Radially outside the TOF system is a 12
radiation length, lead-gas barrel shower counter (BSC) which
measures the energies of electrons and photons over $\sim$80\% of
the total solid angle with an energy resolution of
$\sigma_E/E=$22\%/$\sqrt{E}$ ($E$ in GeV). Outside of the solenoidal
coil, which provides a 0.4~tesla magnetic field over the tracking
volume, is an iron flux return that is instrumented with three
double layers of  counters that identify muons of momentum greater
than 0.5 GeV/$c$.

\subsection{General selection criteria}\label{sub}

All the charged tracks are reconstructed in the MDC, and the
number of charged tracks is required to be four with net charge
zero. Each track should (1) have a good track fit; (2)
have $|\cos\theta|< 0.8$, where $\theta$ is the polar angle of the
track measured by the MDC; (3) originate from the interaction region,
$\sqrt{V_x^2+V_y^2}<2$ cm and $|V_z|<20$ cm, where $V_x$, $V_y$, and
$V_z$ are the x, y, and z coordinates of the point of closest approach
of the track to the beam axis; and (4) be identified as either a pion
or a kaon. The particle identification (PID) is accomplished using the
TOF information and the $dE/dx$ information from the
MDC~\cite{pid}. For instance, a pion should have a higher particle PID
confidence level than those for other hypotheses (kaon, proton).

A neutral cluster in the BSC is considered to be a photon candidate
when its energy deposit in the BSC is greater than 30 MeV, the angle
between the nearest charged track and the cluster is greater than
10$^{\circ}$ (to reject photons associated with the charged
particles), the first hit appears in the first five layers of the BSC
(about six radiation lengths of material), and the angle between the
cluster development direction in the BSC and the photon emission
direction is less than 25$^{\circ}$. If the angle between two photons
is less than 15$^{\circ}$, and the invariant mass of these two photons
is less than 60 $\hbox{MeV}/c^2$, the two photons are merged.

\subsection{Analysis of
        \boldmath{$J/\psi\rightarrow\gamma\phi\rho$}}

For the channel $J/\psi\to \gamma \phi\rho$~$(\phi\to K^+K^-$,
$\rho\to \pi^+\pi^-)$, we require two pions, two kaons, and at least
one photon with energy
greater than 50 MeV. Next, the selected charged
tracks and photon are fitted kinematically using energy and momentum
conservation constraints (4C), looping over all photon candidates. The
combination with the minimum $\chi^2$ is selected, and the photon in
this combination is taken as the radiative photon. Next, we require
the energy of the radiative photon to be greater than 0.1 GeV and the
fit $\chi^2$ to be less than 8.0. The $\chi^2$ requirement is
determined by optimizing the signal ($S$) to noise ($B$) ratio
$(S/\sqrt{S+B})$, where $S+B$ is determined from data and $S$ is
determined from Monte Carlo (MC) simulation after event selection.

In order to suppress backgrounds from $J/\psi$ two body decay channels
with a $\phi$, for example $J/\psi\to \phi\eta$ and $J/\psi\to
\phi\eta^\prime(958)$, the $\phi$ momentum is required to satisfy
$P_\phi \leq$ 1.1 GeV/$c$. To select $\phi$ particles,
the $K^+K^-$ invariant mass is required to satisfy
$|M_{K^+K^-} - M_\phi|\leq0.01$ GeV/$c^2$.

The main expected background channels can be divided into the
following five groups: (1)~$J/\psi\to\phi\eta$,
$\phi\eta^\prime(958)$; (2)~$J/\psi\to \omega KK$, $\phi
KK~(\phi\to\pi^+\pi^-\pi^0)$; (3)~$J/\psi\to\omega
f_0(980)~(f_0(980)\to K^+K^-)$; (4)~$J/\psi\to \phi \pi\pi$, $\phi
f_0(980)$; and (5)~$J/\psi\to (\gamma)
K^\ast(892)\overline{K^\ast}(892)$, $(\gamma,
\pi^0)\pi^+\pi^-K^+K^-$. Since $J/\psi\to \pi^0\rho\phi$ is
forbidden by C-parity conservation, it can be neglected in our
background analysis. The $\pi^+ \pi^-$ invariant mass distributions
from all above possible background channels are smooth with no
$\rho$ peak according to MC simulations; therefore they will not
affect the determination of the number of signal events.

\begin{figure}[hbt]
\centering
\includegraphics[width=6.5cm, height=6cm]{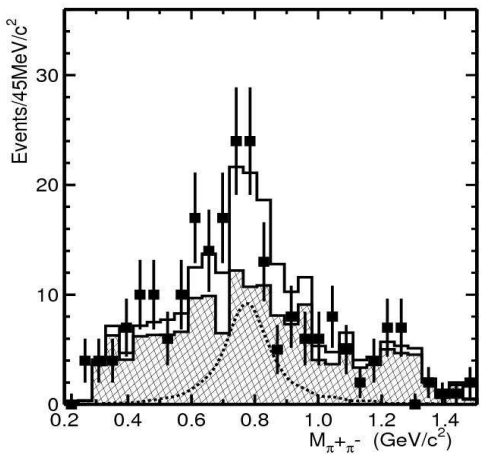}
\caption{The $\pi^+\pi^-$ invariant mass distribution. The square
points with error bars are data, the shaded histogram is from the
$\phi$ sideband events, the dotted curve is the signal shape from
MC simulation, and the blank histogram is the fit.} \label{fig8}
\end{figure}

The $\pi^+\pi^-$ invariant mass distribution for events that survive
the selection criteria is shown in Fig.~\ref{fig8}, and the
$\pi^+\pi^-$ invariant mass distribution from $\phi$ sidebands
($1.05<M_{K^+K^-}<1.08~
\hbox{GeV}/c^2~\hbox{or}~0.985<M_{K^+K^-}<0.992~\hbox{GeV}/c^2$)
scaled to the amount of background in the signal region is also
shown as the shaded histogram.
\par

By fitting the $\pi^+\pi^-$ invariant mass distribution with a
$\rho$ signal shape obtained from MC simulation and using the
histogram from $\phi$ sideband events to describe the background
shape, $43.2\pm 18.8$ signal events are obtained, as shown in
Fig.~\ref{fig8}. The $\rho$ signal statistical significance is
estimated by comparing the likelihood values with and without the
signal in the fit, and it is only about 2$\sigma$. The  detection
efficiency  is $(3.33\pm 0.04)$\% from MC simulation.

\begin{figure}[hbt]
\centering
\includegraphics[width=8.5cm, height=8.5cm]{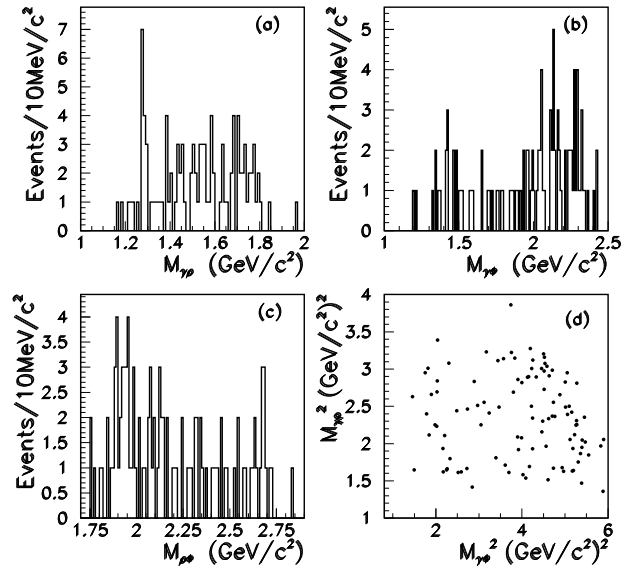}
\caption{(a) The $\gamma\rho$ invariant mass spectrum after $\phi$
selection ($|M_{K^+K^-} - M_\phi|\leq0.01$ GeV/$c^2$), (b) the $\gamma
\phi$ invariant mass spectrum after $\rho$ selection ($|M_{\pi^+\pi^-}
- M_\rho|\leq0.15$ GeV/$c^2$), (c) the $\rho\phi$ invariant mass
spectrum, and (d) the Dalitz plot of $M^2_{\gamma \phi}$ versus
$M^2_{\gamma \rho}$.}
\label{grpgr}
\end{figure}
\par

Figures~\ref{grpgr} (a) and (b) show the invariant mass distributions
of $\gamma \rho$ and $\gamma \phi$; and Figs.~\ref{grpgr} (c) and (d)
show the $\rho \phi$ mass distribution and Dalitz plot of $M^2_{\gamma
  \phi}$ versus $M^2_{\gamma \rho}$ for
$J/\psi\rightarrow\gamma\phi\rho$ candidates, where, the $\pi^+\pi^-$
invariant mass of the $\rho$ candidates must satisfy $|M_{\pi^+\pi^-}-
M_\rho|\leq 0.15 GeV/c^2$. No clear structure around 1440
$\hbox{MeV}/c^2$ region is observed in the $M_{\gamma \rho}$
distribution. There may be some structures in the $\gamma\phi$ and
$\rho\phi$ mass spectra, but because of the low statistics, it is
difficult to determine whether they are real resonances or just
statistical fluctuations.

The systematic errors, are evaluated with selected samples that are
compared with MC simulations.  In this analysis, the systematic
errors on the branching fraction mainly come from the following
sources:

(1). {\it Particle identification (PID).} In Ref.~\cite{pidpik}, the
PID efficiencies of pions and kaons are analyzed in detail. Here, two
charged tracks are required to be identified as pions, and the other two
are required to be kaons, so the systematic error
from PID should be less than 4\%.

(2). {\it MDC tracking and kinematic fit.} In order to study the
systematic errors from the MDC tracking and kinematic fit, many
distributions from data, including the wire efficiency and space
resolution of hits in the MDC, are compared with those from MC
simulations, using two different treatments of the wire resolution
simulation. The difference between the two simulations is taken as
the systematic error for the tracking~\cite{pidpik}. In this paper,
13.5\% is conservatively taken to be the systematic error from MDC
tracking and kinematic fit.

(3). {\it Photon detection efficiency.} The photon detection
efficiency is studied using $J/\psi\to \rho^0\pi^0$ in
Ref.~\cite{pidpik}. The results indicate that the systematic error
is less than 2\% for each photon.

(4). {\it Background uncertainty.} In order to determine the
background uncertainty, several background assumptions were tried,
including: (a) a fourth order polynomial function, (b) the
background shape from MC simulations, and (c) the histogram from
$\phi$ sideband events. The differences between the different fit
results are taken as the systematic error, which is about 13.0\%.

(5). {\it Intermediate decay branching fraction and the uncertainty of
  the number of $J/\psi$ events.} The $\phi$ decay branching fraction
(1.2\%) from Ref.~\cite{pdg06}, and the uncertainty in the total
number of $J/\psi$ events (4.72\%) are also considered as sources of
the systematic error.

Adding all errors in quadrature, the total error is about 20.0\%.

Finally the branching fraction is:
\[
Br(J/\psi\to \gamma\phi\rho)=(4.5\pm 2.0\pm 0.9) \times 10^{-5},
\]
where the first error is statistical and the second is the
systematic.
Since the statistical significance of the $\rho$
signal is only 2$\sigma$, the upper limit (90\% C.L.) is also
determined by a Bayesian method~\cite{pdg06}:
\[
Br(J/\psi\to \gamma\phi\rho) < 8.8 \times 10^{-5}.
\]

\subsection{Analysis of
  \boldmath{$J/\psi\rightarrow\gamma\omega\rho$}}

For the channel $J/\psi\to \gamma\omega\rho ~(\omega\to
\pi^+\pi^-\pi^0, \rho\to \pi^+\pi^-, \pi^0\to \gamma\gamma)$, we
require four pions and greater than two photons, where the
energy of the photon candidates should be greater than 50 MeV. Next,
the selected charged tracks and three photons are fitted using a 4C
kinematic fit, looping over all photon candidates, under the
hypothesis of $J/\psi\to 3\gamma 2(\pi^+\pi^-)$.  The combination with
the minimum $\chi^2$ is selected, and the $\chi^2$ is required to be
less than 9.0, which corresponds to the best signal-noise ratio.  The
two photons with invariant mass closest to the $\pi^0$ mass are
regarded as being from $\pi^0$ decay, and the other is taken as the
radiative photon. Finally, a 5C kinematic fit is made under the
$J/\psi\to \gamma 2(\pi^+\pi^-) \pi^0$ hypothesis with the invariant
mass of the $\gamma \gamma$ pair associated with the $\pi^0$ being
constrained to $m_{\pi^0}$. After the 5C kinematic fit, we require
$\chi^2_{5C}$ to be less than 9.0 and the energy of the radiative
photon to be greater than 0.1 GeV.

The $\pi^+ \pi^-$ and  $\pi^+ \pi^- \pi^0$ combinations for the $\rho$
and $\omega$ candidates are selected from the minimum value of
\par
\[
 \sqrt{
   \left(
          \frac{M_{\pi^+_1\pi^-_2}-M_{\rho}}
                  {\sigma_\rho}
   \right)^2 +
   \left(
          \frac{M_{\pi^+_3\pi^-_4\pi^0}-M_{\omega}}
                  {\sigma_\omega}
   \right)^2},
 \]
\par\noindent
where $\sigma_\rho$ and $\sigma_\omega$ are the
widths (the mass resolutions are included) of $\rho$ and $\omega$,
respectively, from MC simulations. To select the $\omega$ candidates,
we require $|M_{\pi^+_3 \pi^-_4 \pi^0}-M_{\omega}|\leq$ 0.023
GeV/$c^2$.

In order to suppress the backgrounds from $J/\psi$ two body decay
channels with an $\omega$, for example $J/\psi\to \omega
\eta$ and $J/\psi\to \omega \eta^\prime(958)$, the $\omega$ momentum
must satisfy $P_\omega \leq$ 1.2 GeV/$c$. Similarly we also use
$P_\rho\le$ 1.1 GeV/$c$ to suppress the backgrounds from $J/\psi$ two
body decay channels with a $\rho$ resonance.

The possible background channels of
$J/\psi\rightarrow\gamma\omega\rho$ can be divided into the following
four groups: (1)~$J/\psi\to \omega\eta$, $\omega\eta^\prime(958)$,
$\omega \pi^+\pi^-$, $\omega f_0(980)$, $\gamma\omega\omega$,
$b_1(1235)^\pm\pi^\mp$; (2) ~$J/\psi\to a_2(1320)\rho$, ~$\rho\eta$,
$\rho\eta^\prime$, $\gamma\rho\rho$, $\gamma \eta_c~(\eta_c\to
\rho\rho)$; (3) ~$J/\psi\to 2(\pi^+\pi^-)\pi^0$, $\gamma
2(\pi^+\pi^-)$, $\gamma\eta\pi\pi$, $\gamma\eta^\prime$; and
(4)~$J/\psi\to \gamma X ~(X\to \eta\pi\pi$, $K^\ast
\overline{K^\ast})$. Since $J/\psi\to \pi^0\omega\rho$ is forbidden
by C-parity conservation, it can be neglected in our background
analysis. The contamination from all the above possible backgrounds
after event selection is found to be very small according to MC
simulation.

\begin{figure}[hbtp]
\begin{minipage}{65mm}
\centering
\includegraphics[width=7cm]{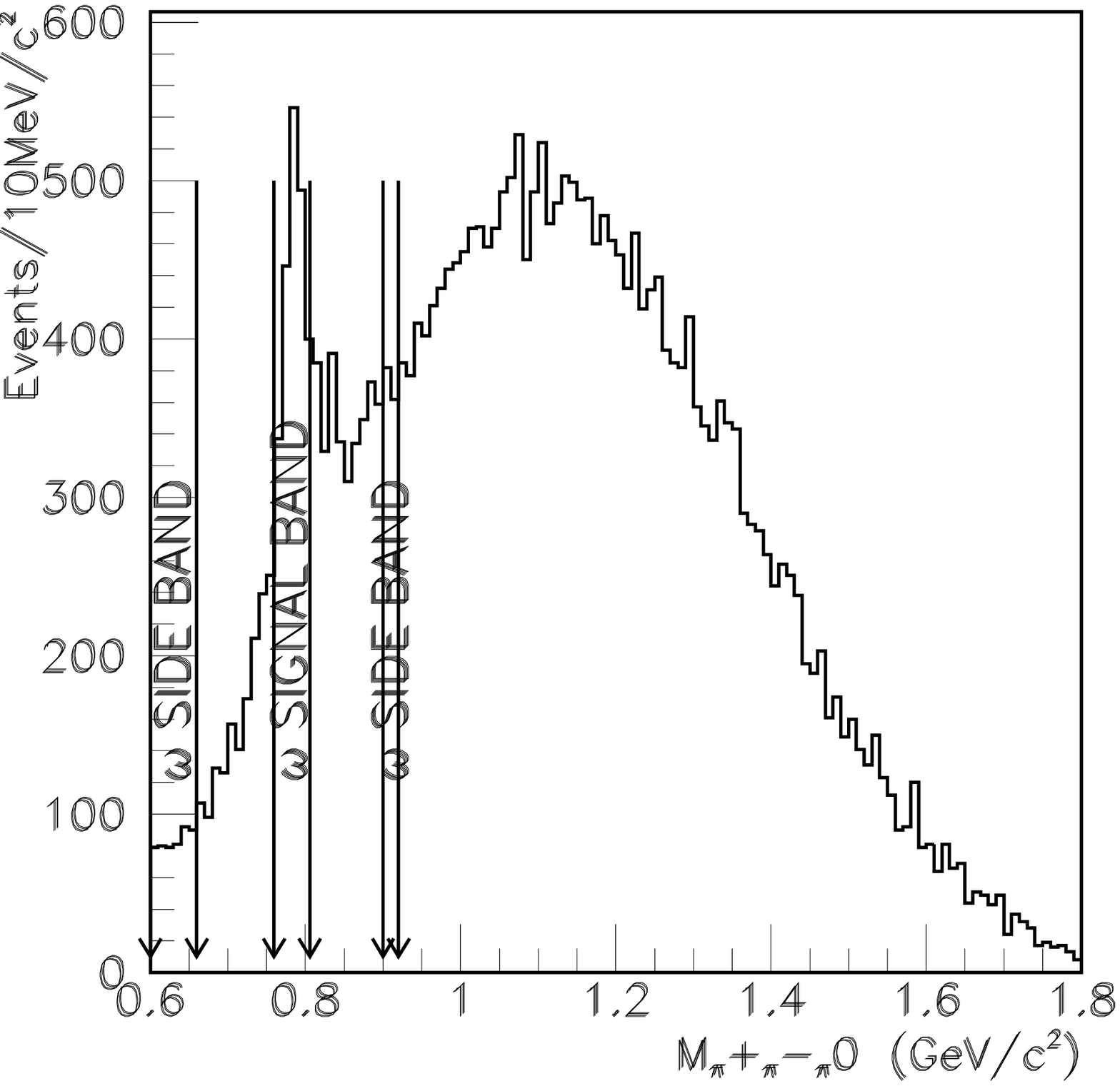}
\caption{The $\pi^+\pi^-\pi^0$ invariant mass spectrum. The $\omega$
signal and $\omega$ side band regions are indicated by arrows.}
\label{fig18}
\end{minipage}
\hspace{\fill}
\begin{minipage}{65mm}
\centering
\includegraphics[width=7cm]{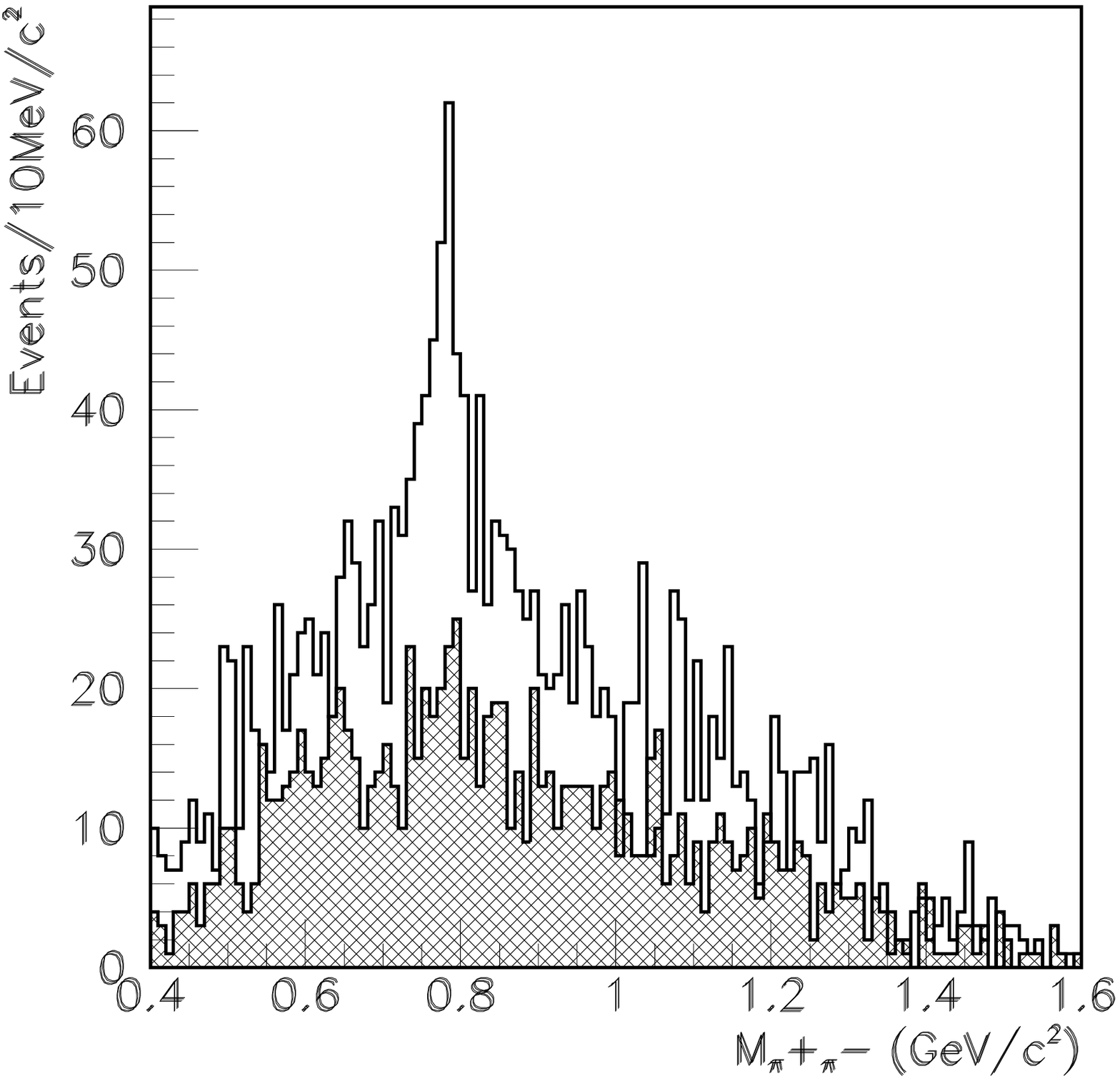}
\caption{The $\pi^+\pi^-$ invariant mass distribution. The shaded
histogram is the distribution from $\omega$ sideband events which has
been scaled to the amount of background in the signal region.} \label{figa}
\end{minipage}
\end{figure}

Figures~\ref{fig18} and \ref{figa} show $\pi^+\pi^-\pi^0$ and
$\pi^+\pi^-$ invariant mass distributions and the $\pi^+\pi^-$
mass distribution of $\omega$ sidebands ($0.6<M_{\pi^+\pi^-\pi^0}<0.67
\hbox{GeV}/c^2~\hbox{or}~0.915<M_{\pi^+\pi^-\pi^0}<0.95
\hbox{GeV}/c^2$), scaled to the amount of background in the signal
region and shown as the shaded histogram in
Fig.~\ref{figa}. After subtracting the $\omega$ sideband events, we
fit the $\pi^+\pi^-$ invariant mass distribution to obtain the number
of $J/\psi\to \gamma\omega\rho$ events.

Although the branching fraction of $\rho\to \pi^+\pi^-$
($\sim$100\%) is about two orders of magnitude larger than that of
$\omega\to \pi^+\pi^-$ (1.7\%), the interference between $\rho$ and
$\omega$ must be considered, and therefore, the fit function is
expressed as~\cite{inter}:
\[
 N(M_{\pi^+\pi^-})=L(M_{\pi^+\pi^-})+
  |A_\rho(M_{\pi^+\pi^-})+A_\omega(M_{\pi^+\pi^-})e^{i\varphi}|^2,
\]
\par\noindent
where $L$ is a polynomial background term, $\varphi$  is the
relative phase angle between the two amplitudes, $A_\rho$ and
$A_\omega$, which are represented by Breit-Wigner functions up to a
numerical factor:
\[
A_V(M_{\pi^+\pi^-})=\sqrt{N_V}F_{BW_V}(M_{\pi^+\pi^-}) (V\equiv
\rho, \omega),
\]
\par\noindent
and the Breit-Wigner functions have an s-independent
width:
\[
F_{BW}=\frac{\Gamma M}{s-M^2+iM\Gamma}.
\]

\begin{figure}[hbtp]
\centering
\includegraphics[width=6.5cm,height=6.0cm]{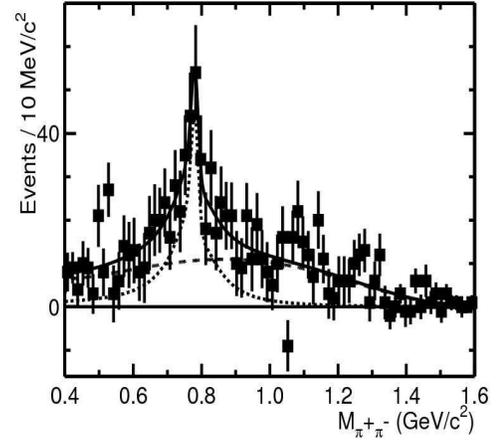}
\caption{The fit,
described in the text, to the $\pi^+\pi^-$ invariant mass
distribution. Here the squares with error bars are data, the
dotted curve is the total signal from $\rho$ and $\omega$ (the
interference between them is also included), the dashed curve is
the polynomial background, and the solid curve is the fit.} \label{fig19}
\end{figure}

Finally, the number of $\rho$ events ($181.3\pm 72.7$), the number
of $\omega$ events ($76.6\pm 61.2$), and the relative phase angle
$(39.4\pm 36.4)^{\circ}$ are obtained; the fit is shown in
Fig.~\ref{fig19}. The  statistical significance of the $\rho$ signal
is about 3.1$\sigma$, and the detection efficiency  is about
$(1.36\pm 0.01)$\% from MC simulation.

\begin{figure}[hbtp]
\centering
\includegraphics[width=8.5cm,height=8.5cm]{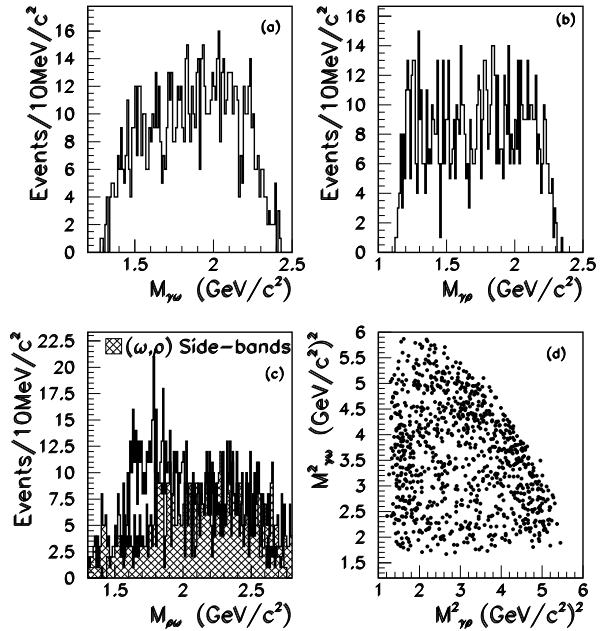}
\caption{(a) The $\gamma\omega$ invariant mass spectrum after $\rho$
selection ($|M_{\pi^+\pi^-} - M_\rho|\leq0.15$ GeV/$c^2$), (b) the
$\gamma \rho$ invariant mass spectrum after $\omega$ selection
($|M_{\pi^+\pi^-\pi^0} - M_\omega|\leq0.023$ GeV/$c^2$), (c) the
$\rho\omega$ invariant mass spectrum (the shaded histogram is the
scaled $\rho\omega$ mass spectrum of $\rho$ and $\omega$ sidebands
events), and (d) the Dalitz plot of $M^2_{\gamma \rho}$ versus
$M^2_{\gamma \omega}$.} \label{grogr}
\end{figure}
\par

Figures~\ref{grogr} (a) and (b) show the invariant mass
distributions of $\gamma \omega$ and $\gamma \rho$, and
Figs.~\ref{grogr} (c) and (d) show the $\rho \omega$ mass
distribution and Dalitz plot of $M^2_{\gamma \rho}$ versus
$M^2_{\gamma \omega}$ for the $J/\psi\rightarrow\gamma\omega\rho$
candidates, where the $\rho$ signal is selected using
$|M_{\pi^+\pi^-} - M_\rho|\leq0.15 GeV/c^2$. No clear structure
around 1440 $\hbox{MeV}/c^2$ region is observed in the $M_{\gamma
\rho}$ distribution.  Although there is a hint of a possible
structure around 1700 $\hbox{MeV}/c^2$ in the $\rho\omega$ mass
spectrum compared with the scaled $\rho\omega$ mass distribution
from $\rho$ and $\omega$ sideband events, it is difficult, because
of the low statistics, to determine whether it is a real resonance
or just due to a statistical fluctuation.  Finally, only the
branching fraction of $J/\psi\to \gamma \omega \rho$ is given.

Systematic errors in the $J/\psi\to \gamma \omega\rho$ branching
fraction measurement are analyzed similarly as in the $J/\psi\to
\gamma\phi\rho$ channel, which mainly come from particle
identification (4\%), the MDC tracking and the kinematic fit
(11.5\%), the photon detection efficiency (6\%), the fitting
procedures and different treatments of the backgrounds (21.6\%), the
total number of $J/\psi$ events (4.72\%), and the $\omega$ decay
branching fraction (0.8\%), taken from Ref.~\cite{pdg06}. The total
systematic error is 26.0\%.

Finally, the branching fraction can be obtained:
\[
 Br(J/\psi\rightarrow\gamma\omega\rho)=(2.6\pm 1.1\pm
0.7)\times 10^{-4},
\]
where the first error is statistical and the second  is the
systematic.
Since the  statistical significance of the $\rho$ signal is only
3.1$\sigma$, the upper limit (90\% C.L.) is also estimated
by a Bayesian method~\cite{pdg06}:
\[
 Br(J/\psi\rightarrow\gamma\omega\rho)<5.4\times 10^{-4}.
\]

\section{Conclusion}
Table~\ref{tab9} summarizes our results for the $J/\psi\to \gamma
VV$ branching fractions; it also lists our $J/\psi\to \gamma
\omega\omega$ branching fraction with one of $\omega$ decays to
$\pi^+\pi^-$, which has a very small branching fraction (15.9\%).
Taking into consideration its large statistical error, this
branching fraction is consistent with the value from
Ref.~\cite{pdg06}.

\begin{table*}
\centering \caption{The branching fractions of $J/\psi\to \gamma
VV$} \label{tab9}
\begin{ruledtabular}
\begin{tabular}{cccc}
    Decay Mode & This Work  & Ref.~\cite{pdg06}
    & BESII$^\dag$ \\ \hline
    $J/\psi\rightarrow\gamma\omega\rho$
    & ($2.6 \pm 1.1 \pm 0.7)\times 10^{-4}$ & & \\
    & ($<5.4\times 10^{-4})$ (90\% C.L.) & &
       \\ \hline
   $J/\psi\rightarrow\gamma\phi\rho$
    & $(4.5 \pm 2.0 \pm 0.9)\times 10^{-5}$ & &\\
    & $(< 8.8\times 10^{-5}$)~(90\% C.L.)  & &
       \\ \hline
    $J/\psi\rightarrow\gamma\omega\omega$
    & $(6.0\pm 4.8\pm 1.8)\times 10^{-3}$ &(1.59 $\pm$ 0.33)$\times 10^{-3}$ &  (2.29 $\pm$ 0.08)$\times 10^{-3}$\\
     one $\omega \to \pi^+\pi^-$ & ($<1.7\times 10^{-2})$~(90\% C.L.)&
     & \\

\end{tabular}
\end{ruledtabular}
\par
$\dag$ Result from
 $J/\psi\rightarrow\gamma\eta(1770)\to \gamma \omega \omega$~\cite{gvv}
\end{table*}

As a check, the $J/\psi \to \gamma\omega\omega$ branching fraction
was fixed at the value from Ref.~\cite{pdg06} and the previous
result from BES Collaboration~\cite{gvv}, and the fit was redone to
obtain the $J/\psi\to \gamma\omega\rho$ branching fraction. This
branching fraction is not sensitive to that of $J/\psi\to
\gamma\omega\omega$.

From our analysis of these two
channels, we did not observe any clear structures in either the $VV$ or
$\gamma V$ mass spectra.

\section{acknowledgments}
The BES collaboration thanks the staff of BEPC and computing center
for their hard efforts. This work is supported in part by the
National Natural Science Foundation of China under contracts Nos.
10491300, 10225524, 10225525, 10425523, 10625524, 10521003, the
Chinese Academy of Sciences under contract No. KJ 95T-03, the 100
Talents Program of CAS under Contract Nos. U-11, U-24, U-25, and the
Knowledge Innovation Project of CAS under Contract Nos. U-602, U-34
(IHEP), the National Natural Science Foundation of China under
Contract No. 10225522 (Tsinghua University), and the Department of
Energy under Contract No.DE-FG02-04ER41291 (U. Hawaii).
\par

\end{document}